\documentclass[prd,12pt,a4paper,nofootinbib,showpacs,tightenlines]{revtex4}
\usepackage[left=2cm,right=2cm,top=2.5cm,bottom=2.5cm]{geometry}

\usepackage[frenchb,english]{babel}
\usepackage[latin1]{inputenc}
\usepackage{enumerate}
\usepackage{amsmath}
\usepackage{amsthm}
\usepackage{amssymb}
\usepackage{graphicx}
\usepackage{floatflt}
\usepackage{fancybox}
\usepackage{stmaryrd}
\usepackage{appendix} 
\usepackage{enumitem}

\usepackage{hyperref} 

\usepackage{color}
\usepackage{caption}

\renewcommand{\Im}{\textrm{Im}}
\newcommand{\Ai}{\textrm{Ai}}

\newcommand{\Om}{\Omega}

\newcommand{\lam}{\lambda}
\newcommand{\Lam}{\Lambda}

\newcommand{\eps}{\epsilon}
\newcommand{\al}{\alpha}

\newcommand{\tp}{{\rm tp}}

\renewcommand{\P}{p_\al}
\newcommand{\eff}{{\rm eff}}
\newcommand{\ma}{{\rm mat}}
\renewcommand{\sc}{{\rm cl}}
\newcommand{\V}{V_{\rm U}}

\newcommand{\p}{\partial}

\newcommand{\Erx}{\mathrm{Er_{h}}}
\newcommand{\Erp}{\mathrm{Er_{p}}}

\newcommand{\be} {\begin{equation}}
\newcommand{\ee} {\end{equation}}
\newcommand{\bsub}{\begin{subequations} \begin{eqnarray}}
\newcommand{\esub}{\end{eqnarray} \end{subequations}}
\newcommand{\bea}{\begin{eqnarray}}
\newcommand{\eea}{\end{eqnarray}}
\newcommand{\bi} {\begin{itemize}}
\newcommand{\ei} {\end{itemize}}
\newcommand{\ben} {\begin{enumerate}}
\newcommand{\een} {\end{enumerate}}
\newcommand{\bmat} {\begin{pmatrix}}
\newcommand{\emat} {\end{pmatrix}} 
\newcommand{\bal} {\begin{aligned}}
\newcommand{\eal} {\end{aligned}}
\newcommand{\btab}{\begin{tabular}}
\newcommand{\etab}{\end{tabular}}

\newcommand{\eq}[1]{Eq.~\eqref{#1}}

\begin{document}
\selectlanguage{english}

\title{Semiclassical momentum representation in quantum cosmology}

\author{Antonin Coutant}
\email{antonin.coutant@aei.mpg.de}
\affiliation{Max Planck Institute for Gravitational Physics, Albert Einstein Institute, Am Muhlenberg 1, 14476 Golm, Germany}

\date{\today}

\begin{abstract}
It is well known that the standard WKB approximation fails to provide semiclassical solutions in the vicinity of turning points. However, turning points arise in many cosmological scenarios. In a previous work, we obtained a new class of semiclassical solutions of the Wheeler-DeWitt equation using the conjugate momentum to the geometric variable. We present here a detailed study of their main properties. We carefully compare them to usual WKB solutions and turning point resolutions using Airy functions. We show that the momentum representation possesses many advantages that are absent in other approaches. In particular, this framework has a key application in tackling the problem of time. It allows us to use curvature as a time variable, and control the corresponding domain of validity, i.e. under which conditions it provides a good clock. We consider several applications, and in particular show how this allows us to obtain semiclassical solutions of the Wheeler-DeWitt equation parametrized by York time.  
\end{abstract}


\maketitle

\newpage 

\section{Introduction}

Despite our lack of a complete theory of quantum gravity, quantum fluctuations of the gravitational field can still be tracked down in some approximate regimes. For this reason, semiclassical approximations play a key role in attempts to characterize quantum gravitational effects, and WKB methods were developed early on in canonical quantum gravity studies~\cite{DeWitt67a}. Such methods are not only useful to obtain corrections to classical general relativity~\cite{Kiefer93}, they are also a crucial tool for tackling conceptual problems such as the problem of time~\cite{Isham92,Kuchar91} or the emergence of a classical background metric~\cite{Banks85,Brout87,Brout88,Parentani98}. Unfortunately, standard WKB techniques fail drastically near a turning point. 

Surprisingly, despite their occurrence in many scenarios, the generic structure of turning points in quantum gravity has not attracted the attention it deserves~\cite{KeskiVakkuri96,Kim99,Kim04}. But turning points are not exotic. They arise in many cosmological models, when the Universe undergoes a bounce~\cite{Bekenstein75,Ashtekar11}, or recollapses due to its spatial curvature~\cite{Kiefer88}, but also in black hole evaporation scenarios such as in 1+1 dimensions~\cite{KeskiVakkuri94,KeskiVakkuri96}. One way to remedy the problem of turning points is to use a different set of variables to build semiclassical solutions. In standard approaches the wave function takes the spatial metric and matter degrees of freedom as arguments. Instead of this \emph{metric representation}, one can work in a \emph{momentum representation}. Since metric variables are canonically conjugated to extrinsic curvature variables, the wave function will now take curvature degrees of freedom as arguments. The idea to use such a momentum representation has recently attracted more attention~\cite{Barvinsky14,Roser14,Roser14b,Coutant14}. While most of these works followed a reduced phase space quantization procedure (the ``identify time before quantization'' of Isham's classification~\cite{Isham92}), our aim is to start from the fully quantized Wheeler-DeWitt constraint and control the validity of the semiclassical momentum representation. In a previous work~\cite{Coutant14} we were concerned with the dynamics of matter fields (``light degrees of freedom'') near a turning point. Here instead, we focus on the properties of the gravitational part (``heavy degrees of freedom''). The objective is to derive the main properties and demonstrate the usefulness of the momentum representation in quantum cosmology. 

We shall compare these solutions of the Wheeler-DeWitt equation, and their validity condition, with standard WKB solutions, and usual resolutions of a turning point singularity. The momentum representation provides much interest compared to alternatives, which we derive and discuss carefully. To start with, it is the \emph{most} semiclassical variable in the vicinity of a turning point, but it is also well defined far from it. As we argue, there is no necessity to ``go back to the geometric representation''. Moreover, it is well-known that an internal variable can play the role of time efficiently only if it behaves semiclassically enough (this has been obtained from a variety of approaches: internal correlations~\cite{Marolf94}, effective techniques~\cite{Bojowald12}, and matter transitions~\cite{Massar97,Massar98}). The semiclassical solutions we study are parametrized by the momentum, and therefore, they allow us to use it as a time variable. We apply this to several examples, and in particular, we show how to apply our framework to the York construction of ``extrinsic time.'' York time played a key role in the demonstration of the well posedness of the Cauchy problem in general relativity~\cite{York72,York73}. It is also a central ingredient in some recent modified gravity theories~\cite{Gomes10}, and was recently used in cosmological settings~\cite{Roser14,Roser14b}.  Here we show that our method exactly allows us to control the approximation necessary for York time to be a good time variable at the quantum level. In the last part of this work, we show that, despite their semiclassical character, these solutions also allow us to access nonclassical quantities, such as tunneling amplitudes. 

In the first section, we review the construction of the (semiclassical) momentum representation, together with the standard (WKB) semiclassical approximation. In the second section, we discuss the key points of our construction, namely its link with the metric representation and the generic structure near a turning point. We then present applications of our framework. In the last section, we show how tunneling amplitudes can be obtained from our framework.

\section{Wheeler-DeWitt equation and momentum representation}
\subsection{Minisuperspace Wheeler-DeWitt equation}
When we restrict ourselves to cosmological systems (homogeneous and isotropic), gravitational degrees of freedom reduce to a single one. The set of possible metrics in this minisuperspace sector is then given by 
\be \label{ds2}
ds^2 = N^2(t) dt^2 - a^2(t) d\Om_K^2, 
\ee
where $d\Om_K^2$ is a fixed spatial metric of constant curvature $K$. The coupled system of matter and gravity is then governed by the minisuperspace action~\footnote{To lighten the notations, we have redefined the gravitational coupling constant $G = \frac{4\pi G_{\rm N}}{3}$, where $G_{\rm N}$ is the standard Newton constant, and $\Lam_{\rm c} = 3 \Lam$ the standard cosmological constant.} 
\be \label{action}
\mathcal S = \frac{\V}{2G} \int \left(- \frac{a \dot a^2}N - \frac{V(a)N}{a} \right) dt + \V \int \left(\sum_j \pi_j \dot{\phi}_j - H_\ma(a; \phi_j) \right)dt , 
\ee
where $H_\ma(a; \phi_j)$ is the Hamiltonian density of matter fields, and $\V$ the volume of the Universe~\footnote{$\V$ appears as a global factor in the classical action \eqref{action}, and therefore, does not affect the classical trajectories. At the quantum level, as we shall see, the volume changes the regime of validity of the semiclassical approximation: the bigger Universe, the more classical it is. In fact, all the error bounds we provide in this work scale as $\V^{-1}$. This fact was used in a different context to advocate for the efficiency of effective equation of loop quantum cosmology~\cite{Rovelli13}.}. $V(a) = -Ka^2+\Lam a^4$ is the Wheeler-DeWitt superpotential. In this setup, it is well known that if the Universe is closed and have a nonzero cosmological constant, there is a turning point~\cite{Kiefer88,Kim99,Kim04}. However, a turning point may arise for a large variety of reasons. It appears when the Universe contains classical exotic matter fields (see e.g.~\cite{Bekenstein75}), or quantum matter with negative energy density (e.g.~\cite{Zeldovich84}) or also in Loop quantum cosmology, where quantum effects forbid the Universe from having a volume smaller than a minimum value~\cite{Ashtekar11}. Our analysis aims at staying rather general, and for this reason, we shall replace the superpotential by an ``effective'' potential $V_\eff(a)$ that contains quantum effects or contributions from fields at equilibrium one wishes to include.

After canonical quantization, the quantum dynamics of the system defined in \eqref{action} is given by a single Hamiltonian constraint: the Wheeler-DeWitt equation, 
\be \label{miniWdW}
\left[\frac{G^2}{a \V^2} \p_a a \p_a + V_\eff(a) + 2Ga \hat H_\ma\right] \Psi(a,\phi_j) = 0.
\ee
Again, to keep the discussion general, we consider a large class of possible matter Hamiltonians, containing all matter degrees of freedom, but also gravitational perturbations~\cite{Halliwell84}. However, for pedagogical purposes, it is instructive to have in mind specific and simple examples. For instance, in a Universe filled with a single scalar field in a potential $V_{\rm scal}$, the matter Hamiltonian reads 
\be \label{Hmat}
\hat H_\ma = \frac{1}{2a^3} \left(-\p_\phi^2 + a^6 V_{\rm scal}(\phi)\right).
\ee 
In this work, we shall make the (drastic) assumption that the Universe is filled with fields at equilibrium. This means that we assume the wave function to factorize as $| \Psi(a) \rangle = \Psi(a) |{\rm mat} \rangle$, where $|{\rm mat} \rangle$ is the state of matter fields in equilibrium (we shall relax this assumption in section \ref{background_Sec}). Doing so, we can replace $\hat H_\ma$ by its mean value $\langle \hat H_\ma\rangle$~\cite{Brout87,Brout88}. For instance, when we consider several massive and massless fields at equilibrium, we have approximately  
\be
\langle \hat H_\ma\rangle = \rho_m + \frac{\rho_r}{a}. 
\ee 
$\rho_m$ is the energy density of nonrelativistic matter and $\rho_r$ that of massless (radiation) fields. This applies at equilibrium, when equipartition is valid. On the contrary, during inflation, under slow roll conditions, the main contribution of the inflaton comes from the potential part, i.e. $\langle \hat H_\ma\rangle = a^3 V_{\rm scal}(\phi)/2$ (and $\phi \sim \mathrm{const}$), giving a cosmological constant contribution~\cite{WeinbergCosmo}. 

In the sequel, we assume a general parametrization of gravitational degrees of freedom. The geometry is governed by a single metric variable $h$, such that $a=a(h)$ in \eq{ds2}~\footnote{Some standard choices would be to use $a$ itself, but also $h = \alpha = \ln(a)$. Choosing a different function $a(h)$ means that one uses different coordinates on the metric superspace. We shall exploit this for instance in Sec.~\ref{York_ex}.}. In terms of $h$, the Wheeler-DeWitt equation takes the very general form 
\be \label{Schro}
\left[ -\p_h^2 + W(h) \right] \Psi(h) = 0.
\ee
For this, we first notice that $\p_a a \p_a = (a_{,h})^{-1}\p_h a (a_{,h})^{-1}\p_h$. Then, we redefine the wave function as
\be \label{Can_det}
\Psi(a) \to \frac{\Psi(h)}{\sqrt{a(h)/a_{,h}(h)}}. 
\ee
By applying this change of wave function to \eq{miniWdW}, we obtain the pseudopotential  
\be \label{Wgener}
W(h) = \underbrace{-a_{,h}^2 \frac{\V^2}{G^2} \Big(V_\eff(a) + 2Ga \langle \hat H_\ma\rangle \Big)}_{W_{\rm cl}(h)} + \underbrace{\V^2 \frac{3 a^2 a_{,h}^{\phantom{1}} a_{,hh}^2 - 2 a^2 a_{,h}^2 a_{,hhh}^{\phantom{1}} - a_{,h}^5}{4G^2 a^2 a_{,h}^3}}_{W_{\rm q}(h)}. 
\ee
We see that the pseudopotential contains two contributions. The first one is the classical pseudopotential $W_{\rm cl}$, written in terms of the general metric variable $h$. The second is the ``quantum pseudopotential'' $W_{\rm q}$. This contribution directly comes from the noncommuting character of $a$ and $\p_a$. While $W_{\rm cl}$ is insensitive to a specific choice of ordering for the Wheeler-DeWitt equation, $W_{\rm q}$ is directly affected. In \eq{miniWdW}, we have chosen the {\it Laplace-Beltrami} ordering. This possesses the interesting property that the Wheeler-DeWitt equation is independent of the way we parametrized the geometry in \eq{ds2}. However, there exist many other legitimate choices for the ordering. Again, for the sake of generality, we shall simply keep $W_{\rm q}$ unspecified, and therefore our results will be valid for any choice, provided that one uses the appropriate $W_{\rm q}$.

\subsection{The momentum-WKB approximation}
\label{pWKB_Sec}
We shall now analyze the Wheeler-DeWitt equation under its very general form \eqref{Schro} so as to build semiclassical solutions. The standard procedure consists of obtaining WKB modes. Those are given by 
\be \label{geoWKB}
\chi(h) = \frac{e^{i \int p_\sc(h') dh'}}{\sqrt{2p_\sc(h)}}.
\ee
We denote the WKB modes by $\chi$ rather than $\Psi$ so as to keep in mind that $\chi$ constitutes an approximate solution. $p_\sc$ is the (classical) conjugate momentum of $h$, and obeys the Hamilton-Jacobi equation
\be \label{hHJ}
p_\sc(h)^2 + W(h) = 0.
\ee
\eq{geoWKB} provides a semiclassical solution of the Wheeler-DeWitt equation \eqref{Schro}, and has been extensively studied in the literature (see~\cite{Isham92,Kiefer93} and references therein). However, \eq{geoWKB} is not an {\it exact} solution. It provides a good approximation when 
\be \label{WKBvalid}
\Erx(h) = \left|\frac{\p_h p_\sc}{p_\sc^2}\right| \ll 1. 
\ee
The error function $\Erx$ gives a bound on the local error due to the WKB approximation~\cite{Winitzki05}. However, when considering a scattering problem, the \emph{accumulated} error might be much smaller, due to destructive interference effects~\cite{Winitzki05,Massar97}. As far as scattering is concerned, the criterion \eqref{WKBvalid} is rather conservative. In order to keep the discussion general, we shall stick to this criterion, which guarantees that the error is \emph{at most} of the order of $\Erx$.  

Unfortunately, the WKB approximation of \eq{geoWKB} breaks down in many physically relevant situations. In particular, it fails dramatically near a turning point~\cite{Isham92,Kiefer13}, where $p_\sc = 0$, or equivalently, $W(h)=0$. Notice that the notion of turning points is classically independent of our choice of metric variable $h$, i.e. of the function $a(h)$. Indeed, we see from \eq{Wgener} that if $W_{\rm cl}$ vanishes at some point, it does for any function $a(h)$, since different $W_{\rm cl}$'s for different $a(h)$'s are proportional. $W_{\rm q}$ on the other hand might affect the location (or presence) of a turning point, although its contribution is often subdominant with respect to $W_{\rm cl}$. 

When switching to the momentum representation, turning points will no longer be singular point of our approximate solutions. In the following, we shall present a slightly generalized version of the framework developed in~\cite{Coutant14}. In the next sections, we will study in detail its properties and some applications. To proceed, we take the Fourier transform of the wave function 
\be \label{FT}
\tilde \Psi(p_h) = \int \Psi(h) e^{-i h p_h} \frac{dh}{\sqrt{2\pi}}.
\ee  
Assuming that this Fourier transform is well defined has nontrivial consequences~\footnote{It is also instructive to notice that \eq{FT} (combined with \eq{Can_det}) is the unitary implementation of the canonical transformation $(\al, p_\al) \to (p_h,h)$, where $\al = \ln(a)$, as is described in~\cite{Barvinsky14} in a similar context.}. First, we assumed that we integrate over all real values of $h$. This creates difficulties if one wishes to use $a$ itself as the metric variable (we shall return to that point in Sec.~\ref{azero_Sec}). Second, \eq{FT} implies that we discard the growing mode in the classically forbidden side of the turning point. As shown in~\cite{Barvinsky14}, a well-defined momentum representation necessarily implies such a selection rule. It is also necessary to recover the background field approximation near a turning point, see the discussion in Sec.~\ref{background_Sec}. After the Fourier transform, the wave function obeys the Wheeler-DeWitt equation in the momentum representation, i.e. 
\be \label{pSchro}
\left[ p_h^2 + W(i\p_{p_h}) \right] \tilde \Psi(p_h) = 0.
\ee
To obtain semiclassical solutions for this equation, we assume a semiclassical \emph{ansatz} 
\be \label{WKBansatz}
\tilde \Psi(p_h) = A(p_h) \exp\left({-i \int^{p_h} h_\sc(p) dp}\right).
\ee
The key point is that this expression gives a good approximation when the amplitude is a slowly varying function of $p_h$. Therefore, we shall solve \eq{pSchro} using a gradient expansion. The phase of \eq{WKBansatz} is the classical action, and is obtained from the Hamilton-Jacobi equation 
\be \label{pHJ}
p_h^2 + W(h_\sc(p_h)) = 0.
\ee
This is the same equation as \eqref{hHJ}, except that we now solve for $h$ as a function of $p_h$ instead of the converse. If the pseudopotential is monotonic, the solution $h_\sc(p_h)$ is unique. If $W(h)$ vanishes, say at $h=h_\tp$, there is a turning point. The region where $W(h)$ is positive (without loss of generality, we assume it for $h < h_\tp$) is classically forbidden. Hence, when $p_h$ runs from $-\infty$ to $+\infty$, $h_\sc(p_h)$ goes from $+\infty$ to $h_\tp$ and back again. 

Before using any approximation, the \emph{ansatz} \eqref{WKBansatz} simply defines the amplitude $A(p_h)$. When applied to \eq{pSchro}, it gives  
\be
\left[p_h^2 + W\Big(h_\sc(p_h) + i\p_{p_h}\Big)\right] A(p_h) = 0. \label{unexpandWKB}
\ee
Since the amplitude is assumed to vary slowly, we expand the operator in \eq{unexpandWKB} in powers of $i\p_{p_h}$. For this, as we detail in Appendix~\ref{pWKB_App}, one must use a Taylor expansion of functions of non-commuting arguments. Here, the first order expansion reads 
\be
W\Big(h_\sc(p_h) + i\p_{p_h}\Big) = W(h_\sc(p_h)) + W'(h_\sc(p_h)) i\p_{p_h} + \frac i2 W''(h_\sc(p_h)) h_\sc'(p_h) + O\left(\p_{p_h}^2 \right). 
\ee
(By convention, $W' = \p_h W$ and $h_\sc' = \p_{p_h} h_\sc$.) Using this, \eq{unexpandWKB} becomes
\be
\left[p_h^2 +W(h_\sc(p_h)) + W'(h_\sc(p_h)) i\p_{p_h} + \frac i2 W''(h_\sc(p_h)) h_\sc'(p_h) \right] A(p_h) = 0. \label{WKBexpan}
\ee
When sorting these terms in gradients, we see that the first order one is nothing other than the classical Hamilton-Jacobi equation \eqref{pHJ}. The second order part of \eq{WKBexpan} gives the amplitude 
\be
A(p_h) = \left|W'(h_\sc(p_h))\right|^{-1/2}. \label{slowA} 
\ee
We then deduce the semiclassical solution of the Wheeler-DeWitt equation 
\be
\tilde \chi(p_h) = \frac{e^{-i \int h_\sc(p_h')dp_h'}}{\sqrt{|W'|}}. \label{pWKB} 
\ee
Postulating \eqref{pWKB} as an exact solution corresponds to the reduced phase space quantization approach~\cite{Barvinsky14}, or ``identify time before quantization''~\cite{Isham92}. In other words, this is what one obtains if one first solves the Hamilton-Jacobi equation for $h_\sc$ and then quantizes. This is not what we obtain starting from the Wheeler-DeWitt equation \eqref{miniWdW}. As we show in appendix \ref{pWKB_App}, \eqref{pWKB} is a good approximation if we have
\be \label{pWKBvalid}
\Erp(p_h) = \left|\frac{W''^2 h_\sc'(p_h)}{W'^2}\right| \ll 1 , 
\ee
where $W$ and its derivatives are evaluated at $h = h_\sc(p_h)$. It is remarkable that this condition cannot be guessed from the somewhat naive statement that the WKB approximation is valid when the ``phase varies much more slowly than the amplitude,'' which would lead to an analog of \eq{WKBvalid} like $|W''/(h_\sc W')| \ll 1$. In addition, \eq{pWKBvalid} shows that the error vanishes on the turning point. Of course a statement like ``momentum WKB is exact on a turning point'' has no meaning; the error must be small for a sufficiently large interval. But in close vicinity of the turning point, momentum WKB is always a good approximation. 
\medskip

In the above construction, the prefix ``semi'' of ``semiclassical'' should not be taken lightly. Indeed, semiclassical solutions of \eq{pWKB} still encode many quantum features of the gravitational field. In particular, no classical background metric exists at the level of \eq{pWKB}. In fact, the validity of our construction, governed by \eq{pWKBvalid}, is independent of internal (i.e., matter fields) degrees of freedom. As we shall now see, this is not the case when considering the classical background limit. In order to recover a classical space-time, we shall use matter fields as probes. They will experience a classical background as long as \emph{energy changes} are small.

\subsection{The background as perceived by matter fields}
\label{background_Sec}

To physically interpret the semiclassical solutions obtained in \eq{pWKB}, we use a matter field as a probe. In other words we ask, how do matter fields perceive the wave function of the gravitational degrees of freedom? For this we consider a massive field, i.e. $V_{\rm scal}(\phi) = M \phi^2$ in \eq{Hmat}. To simplify the discussion, we assume that this field evolves {\it adiabatically}, that is, no particle production occurs while the Universe expands. This might be a bad approximation in realistic scenarios. However, it drastically simplifies the discussion, without altering the main conclusions. In~\cite{Coutant14}, we considered particle production due to interactions and the conclusions we shall draw are maintained. In the adiabatic limit, matter states are adequately described by their decomposition in the adiabatic Fock basis~\cite{Fulling,BirrellDavies}. To build this basis, we consider the instantaneous eigenvectors of the matter Hamiltonian 
\be
\hat H_\ma |n \rangle = M n |n\rangle. 
\ee
The quantum number $n$ is conserved in the adiabatic limit. Hence, the Wheeler-DeWitt equation \eqref{miniWdW} decouples into several second order ordinary differential equations, one for each value of $n$. We then apply all the preceding results using several pseudopotentials 
\be
W_n(h) = -a_{,h}^2 \frac{\V^2}{G^2}\Big(V(a) + 2Ga M n \Big) + W_{\rm q}(h). 
\ee
For each pseudopotential, we construct the semiclassical solution $\tilde \chi_n(p_h)$ using \eq{pWKB} and the corresponding Hamilton-Jacobi solution $h_n(p_h)$. We then obtain the general semiclassical solution of the Wheeler-DeWitt equation as  
\be \label{multiWKB}
\tilde \Psi(p_h) = \sum_n c_n\frac{e^{-i \int h_n(p_h')dp_h'}}{\sqrt{|W_n'|}} |n\rangle,
\ee
where $c_n$'s are constants. Such a solution is an arbitrary superposition of matter eigenstates associated with a semiclassical state for the gravitational degrees of freedom. When adopting the point of view of matter, the interpretation of this solution is rather clear. The semiclassical wave function for gravity is analogous to $e^{- i E_n t}$, i.e. it describes the ``time evolution,'' where $p_h$ plays the role of time. From this point of view, there are several crucial advantages with respect to standard discussions of the WKB interpretation of quantum cosmology~\cite{Isham92,Kiefer93,Bertoni96}. First, since we obtained semiclassical solutions in the momentum representation, they are perfectly adequate to describe physics near a turning point (as shown by \eq{pWKBvalid}). Second, we can consider not only one semiclassical solution, but a superposition of several of them. In other words, \eq{multiWKB} does not describe matter in a single background. Each matter state perceives its own background. Third, one is not restricted to consider one matter eigenstate, but {\it any} superposition of eigenstates. In the presence of interactions or non-adiabaticities, quantum transitions occur between these matter states. Quantum transitions were discussed in~\cite{Massar97,Massar98} in the metric representation and in~\cite{Coutant14} in the momentum representation. The notion of a single background metric emerges from \eq{multiWKB} only when the spread in matter energy is small, i.e., $\Delta E \ll \bar E$. As in~\cite{Massar97,Massar98,Coutant14}, we consider energy changes at first order around its mean value. We then have 
\be \label{BFA}
h_n(p_h) \simeq \bar h(p_h) + \p_E h(p_h) (E_n - \bar E), 
\ee
where $\bar h(p_h)$ is the solution of the Hamilton-Jacobi equation \eqref{pHJ} for $E = \bar E$. The first factor in the second term gives rise to the background notion of \emph{time} via the Hamilton-Jacobi relation 
\be \label{WKBtime}
t(p_h) = \int \p_E h(p_h') dp_h'. 
\ee
This is exactly what is needed to change the phase of \eq{multiWKB} from $\int dp_h'$ into $\int dt'$. Therefore the solution \eqref{multiWKB} becomes 
\be \label{backWKB}
\tilde \Psi(p_h) = \frac{e^{-i \int \bar h(p_h')dp_h' + i \int \bar E(t') dt'}}{\sqrt{|W_{\bar n}'|}} \sum_n c_n e^{- \int E_n(t') dt'} |n\rangle. 
\ee
This now corresponds to a superposition of matter eigenstates living in a background metric characterized by $\bar p_h(t)$ from \eq{WKBtime}, where the geometry is sourced by the mean value of matter energy $\bar E$~\cite{Brout87}. The time $t$ that emerges from \eq{WKBtime} coincides with the ``WKB time''~\cite{Isham92,Kiefer93,Bertoni96} and our procedure is perfectly valid near a turning point. Indeed, we recognize in \eq{backWKB} the standard from of a Born-Oppenheimer approximation, where $\tilde \Psi = \tilde \chi(p_h) \psi(\phi; p_h)$, i.e. the wave function of the heavy part factorizes. A key aspect of this approach is that $t$ is really the time as perceived by matter fields. It arises as the conjugate momentum of matter energy~\footnote{Of course, even in a fixed background, there are many choices of a time coordinate. The choice comes here from the identification of matter energy in the Hamiltonian contraint. By defining for instance $a\hat H_\ma$ instead of $\hat H_\ma$, one obtains the conformal time instead of the co-moving time~\cite{Parentani96,Coutant14}.}. We underline once more that no background time (or metric) exists at the level of \eq{multiWKB}. Hence using $p_h$ as a time variable is more fundamental than using the WKB time $t$ of \eqref{WKBtime}. However, one still needs to consider a semiclassical approximation to have a sensible notion of evolution, as a phase for each matter state. This is even clearer when considering quantum transitions, as \eq{multiWKB} induces a \emph{unitary} evolution for these transitions. In~\cite{Coutant14} it was obtained that the evolution is unitary when the momentum WKB approximation is valid and the various $h_n(p_h)$'s describing semiclassical trajectories have the same monotonicity in $p_h$. 

At this point, we emphasize that we do not claim that $p_h$ is a \emph{fundamentally} better variable, or ``time''. Very similar conclusions to that drawn in Sec.~\ref{background_Sec} were obtained using WKB methods in metric representation~\cite{Banks85,Brout88,Kiefer93,Bertoni96,Massar98}. However, the validity conditions in both representations are quite different, as shown by comparing Eqs.~\eqref{WKBvalid} and~\eqref{pWKBvalid}. In particular, close to a turning point, $p_h$ is the most semiclassical variable, and hence gives the best time variable. 

We now discuss the validity of the background field approximation, i.e., of \eq{BFA}. Since the matter Hamiltonian appears linearly in the total Hamiltonian constraint, so does the matter energy in the pseudopotential of \eq{Wgener}. Using the definition of the Hamilton-Jacobi solution $h_\sc$, we compute the condition for the first order expansion in the energy change $\Delta E$: 
\be \label{BFAvalid}
\frac{\Delta E \p_E^2 h_\sc}{\p_E h_\sc} = \frac{\Delta E \left|W'' \p_E W - W' (\p_E W)' \right|}{W'^2} \ll 1.
\ee
Unlike \eq{pWKBvalid}, this condition depends on the matter degrees of freedom through the energy fluctuations $\Delta E$. In a large Universe, the total inertia of gravity, governed by $\bar E$ is much larger than energy changes, since the former is proportional to the total mass of all the particles and the latter of the mass of a single particle. Therefore, there is generally a hierarchy between the two approximations encoded in \eq{pWKBvalid} and \eq{BFA}. The semiclassical approximation is in general much better than the (more drastic) classical background approximation. This means that the solutions of \eq{pWKB} are not only useful to recover quantum field theory in curved space-time, but they also encode some quantum corrections. 

\section{metric representation and boundary conditions}
\label{BC_Sec}

\subsection{Inverse Fourier transform at the saddle point approximation \label{invFT_Sec}}
Having obtained semiclassical solutions in the momentum representation, it is instructive to relate them to semiclassical solutions in the metric representation. For this, we start from $\tilde \chi(p_h)$ of \eq{pWKB} and compute its Fourier transform 
\be \label{inverseFT}
\Psi(h) = \int \frac{e^{-i \int h_\sc(p_h')dp_h' + i h p_h}}{\sqrt{2\pi |W'|}} dp_h. 
\ee
This integral gives a (semiclassical) solution in the metric representation, but one whose validity is controlled by \eq{pWKBvalid} rather than \eq{WKBvalid}. In particular, it is valid across a turning point. As we shall now see, when \emph{both} \eqref{pWKBvalid} and \eqref{WKBvalid} are valid, this integral coincides with usual WKB solutions as in \eq{geoWKB}. To show this, we evaluate the integral using the {\it saddle point method}~\cite{Olver}. This method precisely requires \eqref{WKBvalid} to be accurate. At each value of $h$, the value of the saddle point $p_h = p_*(h)$ solves the equation 
\be \label{spHJ}
h = h_\sc(p_*(h)). 
\ee
Hence, $p_*(h)$ is nothing other than the classical momentum function $p_*(h) = p_\sc(h)$, solution of the Hamilton-Jacobi equation \eqref{hHJ}. Moreover, the phase evaluated at the saddle point gives the \emph{Legendre transform} of the action $\int h_\sc(p_h')dp_h'$; that is, the action in the metric representation. Indeed, from a change of integration variable $p_h' = p_\sc(h')$ and an integration by parts follow the identity 
\be
-\int^{p_\sc(h)} h_\sc(p_h')dp_h' + h p_\sc(h) = \int^h p_\sc(h') dh' . 
\ee
The result of the integration \eqref{inverseFT} at the saddle point approximation then gives 
\be \label{spFT}
\Psi(h) = \frac{e^{i \int^h p_\sc(h') dh'}}{\sqrt{|h_\sc' W'(h)|}}, 
\ee
where the irrelevant global phase has been discarded. Moreover, deriving \eq{pHJ} with respect to $p_h$ shows that $h_\sc' W' = -2p_h$. Therefore, the prefactor arising from the saddle point approximation is exactly what is needed to obtain the WKB amplitude in metric representation and \eq{spFT} simply becomes $\Psi(h) = \chi(h)$. 
This is a particular case of the fact that different Dirac represenations that are classically related by a canonical transformation are semiclassically unitarily equivalent~\cite{Barvinsky96,Barvinsky14}. Our method allows us to control the validity of this equivalence, which requires both \eq{WKBvalid} and \eqref{pWKBvalid}. 

When there is a turning point, there are two solutions of the Hamilton-Jacobi equation \eqref{spHJ} and therefore one must sum over these two saddle point contributions. Moreover, the Hamilton-Jacobi equation \eqref{pHJ}, equivalent to \eqref{spHJ}, is invariant under $p_\sc \to - p_\sc$. Hence if $p_\sc(h)$ is one solution, the other is $-p_\sc(h)$. The result for the wave function becomes 
\be \label{spReflection}
\Psi(h) = \frac{e^{i \int^h p_\sc(h') dh' - i \frac\pi4}}{\sqrt{|2p_\sc(h)|}} + \frac{e^{-i \int^h p_\sc(h') dh' + i \frac\pi4}}{\sqrt{|2p_\sc(h)|}}. 
\ee
One solution represents a contracting Universe, the other an expanding one. Of course, one can only obtain this expression far away from the turning point. Indeed, the validity condition for the saddle point approximation basically reduces to \eq{WKBvalid}, which is valid away from the turning point. 

\subsection{Behavior of the wave function near the singularity $a=0$}
\label{azero_Sec}

A standard choice to parametrize the metric is $\al = \ln(a)$~\cite{Kiefer13}, which presents several advantages. First, the quantum potential of \eq{Wgener} vanishes, and second, while the wave function is only defined for positive values of $a$, $\al$ runs from $-\infty$ to $+\infty$. However, one could be willing to work directly with the scale factor $a$, as it allows several simplifications (such as the superpotential becoming polynomial). Classically, the scale factor is restricted to $a > 0$. Quantum mechanically, it is an open issue whether one would like to keep this constraint or not~\cite{Isham92}. This question is the minisuperspace manifestation of a more general issue in quantum gravity, and is referred to as the \emph{spatial metric reconstruction problem}~\cite{Isham92}. When constructing the momentum representation in \eq{FT}, we had to consider all real values of $a$ and this question appears to be important for our construction. 

The main problem of considering negative values of $a$ is an interpretational one. What meaning should one give to $a < 0$? As emphasized in Sec.~\ref{background_Sec}, we adopt here the point of view of matter. In particular, we are not trying to answer a question such as ``what is the probability of $a$ having such value?'' Hence, $a$ should not necessarily be interpreted as part of a metric at the fundamental level, but only in the regime where matter fields approximately propagate as in a classical background. Since we considered scenarios with a turning point, we always have $a>0$ in the semiclassical limit. To further support this, we evaluate the semiclassical wave function obtained in the momentum representation at $a=0$. From \eq{inverseFT}, it follows that 
\be 
\Psi_{a=0} = \int \frac{e^{-i \int a_\sc(p')dp'}}{\sqrt{2\pi |W'|}} dp. 
\ee
Because $a=0$ is within the classically forbidden region, the saddle point $p_{*}$ is \emph{complex}. As a consequence, the value of the wave function is exponentially small: 
\be
|\Psi_{a=0}|^2 \sim \frac{e^{- \Im \int^{p_*} a_\sc(p')dp'}}{2\pi |W_*'|}. 
\ee
This means that whether we want to impose that $\Psi$ only has support on $a>0$ or not, the difference between the two choices is exponentially suppressed. Of course, the above equation is valid only if the turning point lies ``far enough'' from $a=0$, more precisely, if \eq{WKBvalid} is valid near $a=0$. If this is not the case, then imposing boundary conditions such as $\Psi_{a=0} = 0$ will deform the wave function and make the momentum semiclassical solution of \eq{pWKB} a bad approximation. In such a case, the most reasonable method is presumably to use a different metric variable such as $\al$.

\subsection{Turning point vicinity}
In a small neighborhood of a bounce, the pseudopotential is well approximated by a linear function 
\be
W(h) = -\kappa (h - h_\tp), 
\ee
where the bounce occurs at $h = h_\tp$. With this pseudopotential, the Wheeler-DeWitt equation \eqref{pSchro} becomes 
\be \label{pAiry}
\left[ p_h^2 -i \kappa \p_{p_h} + \kappa h_\tp \right] \tilde \Psi(p_h) = 0.
\ee
Since it involves only a first order derivative in $p_h$, the momentum-WKB approximation is {exact}. The solution of \eq{pAiry} is given by 
\be \label{pWKBAiry}
\tilde \Psi(p_h) = \kappa^{-1/2} \exp\left({-i (p_h^3/(3\kappa) + h_\tp p_h)}\right),
\ee
where the normalization constant is chosen to be the same as in \eq{pWKB}. It is now instructive to use the procedure of Sec.~\ref{BC_Sec} and relate this solution to WKB solutions in metric representation. From \eq{inverseFT}, we write the wave function in the metric representation as 
\be
\Psi(h) = \int \exp\left({-i (p_h^3/(3\kappa) + h_\tp p_h) + i p_h h}\right) \frac{dp_h}{\sqrt{2\pi \kappa}}.
\ee
This is the integral representation of an Airy function~\cite{Olver}~\footnote{It is noticeable that in~\cite{Coutant14}, we also found that the \emph{matter part} of the wave function is described by an Airy function in the vicinity of the turning point. However, the two Airy functions are very different. Here, it describes the wave function of the geometric degrees of freedom, while in~\cite{Coutant14}, it described the probability amplitude of creating particles close to a turning point. In particular, both define a region of ``close vicinity of the turning point'' (see the discussion after \eq{asAiry} here and after (44) in~\cite{Coutant14}), but the characteristic sizes are different.}. Hence, 
\be \label{Airy}
\Psi(h) = \sqrt{\frac{2\pi}{\kappa^{1/3}}} \Ai\left(-\kappa^{1/3}(h - h_\tp)\right).
\ee
On the right side ($h-h_\tp > 0$), when going far from the turning point, the Airy function reduces to a sum of oscillatory terms: 
\be \label{asAiry}
\Psi(h) \sim \frac{e^{i \frac23 \kappa^{2}|h - h_\tp|^{3/2} - i \frac{\pi}4}}{\sqrt{2\kappa^{1/2}|h - h_\tp|^{1/2}}} + \frac{e^{-i \frac23 \kappa^{2}|h - h_\tp|^{3/2} + i \frac{\pi}4}}{\sqrt{2\kappa^{1/2}|h - h_\tp|^{1/2}}}.
\ee
We recognize the sum of two semiclassical solutions, obtained from \eq{geoWKB}. As in \eq{spReflection}, one describes a contracting Universe and the other an expanding one. On the other side of the turning point, for $h - h_\tp < 0$, the wave function decays exponentially. This corresponds to the classically forbidden region, i.e., the Universe undergoes a bounce and hence never reaches very small densities. Note that in the scenario of a recollapsing Universe, the discussion above is the same {modulo} the replacement $W \to -W$, that is, the classically forbidden region lies at large volumes. What we learn here is that close to the turning point, the wave function cannot be approximated by semiclassical solutions in the metric representation. Far from the turning point, when $|h - h_\tp| \gtrsim \kappa^{1/3}$, \eq{asAiry} becomes valid and one can interpret the solution as a superposition of two semiclassical solutions. ($\kappa^{1/3}$ corresponds to the ``region of validity of WKB'' of~\cite{KeskiVakkuri96}.~\footnote{Our solution \eqref{Airy} corresponds to the function $v_\lambda$ of reference~\cite{KeskiVakkuri96}, and described in their section 2.3, but it was not identified as an Airy function.}) 

The resolution of the singular character of the WKB approximation near a turning point with an Airy function, as in \eq{Airy}, is a standard method, which can be generalized using the Green-Liouville approach~\cite{Kim99,Kim04}. We showed here how the momentum representation is directly related to it. However, in many cases, it is advantageous to \emph{stay in momentum representation}, and not make use of Airy functions or the geometric representation. First, the momentum representation is not restricted to a linear potential, as shown in Sec.~\ref{pWKB_Sec}, unlike \eq{Airy}. Second, the solution in the momentum representation \eqref{pWKBAiry} is valid everywhere, unlike \eq{asAiry}. As discussed in Sec.~\ref{background_Sec} one can then use this representation to follow the evolution of matter states all along, i.e., use $p_h$ as a time variable. This comparison with the Airy function also illustrates the limitations of the proposal to extend the WKB notion of time beyond the condition \eqref{WKBvalid}. In this approach~\cite{Padmanabhan90}, one uses the phase of the wave function as a time. But \eq{Airy} is purely real, and no such construction is possible in this case. On the contrary, in the momentum representation, the wave function is semiclassical and $p_h$ is a perfectly viable time variable.

\section{Applications}
\label{app_Sec}

\subsection{An exactly solvable example}
\label{Solvable_ex}
We now apply the preceding results in several simple examples. This illustrates how the procedure is implemented, and what are the ingredients that make the semiclassical solution \eqref{pWKB} valid. We consider a Universe with a positive cosmological constant and filled with non relativistic {\it quantum} matter at equilibrium. The energy density of the quantum matter is assumed to be {\it negative}. This induces a repulsive gravitational force that will generate a bounce. Using the metric variable $h = \al = \ln(a)$, the potential then becomes 
\be \label{toyW}
W(\al) = - \frac{\V^2}{G^2} \Big(\Lam e^{6\al} - 2G \rho_q e^{3\al} \Big),
\ee 
where $\rho_q >0$ so that $-\rho_q$ is the energy density of quantum matter. To build semiclassical solutions, we first solve the Hamilton-Jacobi equation \eqref{pHJ}. We obtain 
\be
\al_\sc(\P) = \frac13 \ln\left(\frac{G \rho_q}{\Lam} + \frac{G \rho_q}{\Lam} \sqrt{ 1 + \frac{\Lam \P^2}{V_{\rm U}^2 \rho_q^2}}\right). 
\ee
We see that $\al_\sc$ decreases from $+\infty$ to $\al_\tp = \ln\left(2G \rho_q/\Lam\right)/3$ where the bounce occurs, and then increases back to $+\infty$. We then directly obtain the semiclassical solution of the Wheeler-DeWitt equation using our momentum-WKB expression \eqref{pWKB}. Note that with the pseudopotential \eqref{toyW}, the Hamilton-Jacobi equation \eqref{pHJ} possesses an infinite number of solutions. However, we easily see from the variations of $W$ that there is a unique solution such that $\al_\sc(\P)$ is {\it real}. Similarly, at the quantum level, \eq{pSchro} possesses derivatives in $\P$ at all orders. However, by using a semiclassical treatment, we basically neglect all other solutions except the one corresponding to the unique real solution of the Hamilton-Jacobi equation \eqref{pHJ}. 

\begin{figure}[!ht]
\begin{center}
\includegraphics[width=0.8\columnwidth]{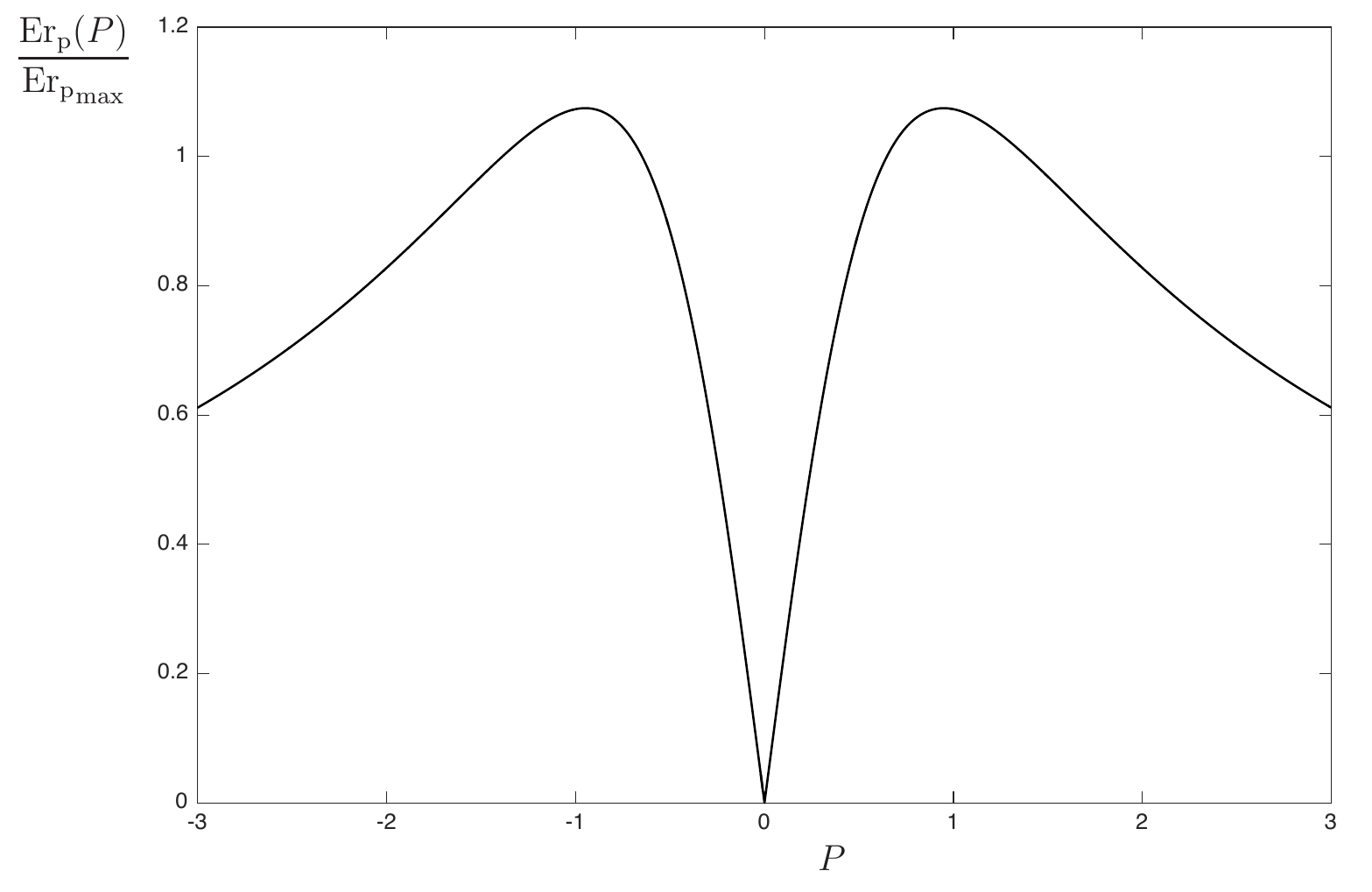}
\end{center}
\caption{Plot of the error function $\Erp/\Erp_{\rm max}$ as a function of $P = \frac{\Lam^{1/2} \P}{\V \rho_q}$.  
}
\label{Error_plot} 
\end{figure}

We now discuss the validity of the semiclassical approximation. For this we use the error function of \eq{pWKBvalid} with the pseudopotential \eqref{toyW}. It is easy to see that $\Erp(p_h) \to 0$ for both $\P \to 0$ and $\P \to \infty$. Therefore, the semiclassical solution \eq{pWKB} is a very good approximation both near the bounce and far from it. The error thus grows from the vicinity of the bounce to a maximum value and then decreases to zero. A quick computation shows that the maximum error is, up to a numerical factor of order 1, given by  
\be
\Erp_{\rm max} = \frac{\Lam^{1/2}}{\V \rho_q},
\ee
see Fig.~\ref{Error_plot}. On can also consider the fluctuations of matter energy $\delta \rho$ around the mean value $\rho_q$. Using \eq{BFAvalid}, we see that the background field approximation is valid all along if 
\be
\frac{\delta \rho \Lam}{\rho_q} \ll 1.
\ee
We point out that, modulo the discussion of Sec.~\ref{azero_Sec}, the same error bounds are obtained when using $a$ instead of $\al$ as a geometric variable. 

\subsection{Cosmology with York time}
\label{York_ex}
In canonical general relativity, one can define a ``canonical '' notion of time using the trace of the extrinsic curvature~\cite{York72,York73}. The obtained time parameter, known as York time, is the conjugate momentum to the volume element $\sqrt{g}$. In minisuperspace, $\sqrt{g} = a^{1/3}$, and hence we apply our results using $h = a^3$, $p_h$ giving York time. Indeed, using the action \eq{action} written in terms of $h = a^3$, one finds 
\be
p_h = -\frac{\V \dot a}{3 G N a}, 
\ee
and therefore, $p_h = \V T$. From the results of the preceding sections, this means that at the quantum level, York time becomes a good time variable only if \eq{pWKBvalid} is satisfied. From \eq{Wgener}, we deduce the pseudopotential 
\be 
W(h) = - \frac{\V^2}{9G^2} \left(\Lam - \frac{K}{h^{2/3}} + \frac{2G \rho_m}{h} + \frac{2G \rho_r}{h^{4/3}} \right) -\frac{\V^2}{4G^2 h^2}. 
\ee 
A peculiarity of cosmology with York time is that the pseudopotential becomes flat for large $h$. Hence, $T$ will vary slowly and one expects the semiclassical approximation \eqref{pWKB} to become inaccurate. This can be confirmed using the validity condition of \eq{pWKBvalid}. York time might be a good time variable to follow the evolution of the state through the turning point, but it will eventually become a bad time variable. On the contrary, the volume, or $h$, will asymptotically be better and better, since WKB in the metric representation becomes exact for $h\to \infty$. Therefore, an accurate description will in this case involve a mix between representations. To illustrate this, we investigate again the preceding example, i.e. a cosmological constant and $\rho_m = -\rho_q$ (and neglecting $W_{\rm q} \propto h^{-2}$). Once again, the Hamilton-Jacobi equation \eqref{pHJ} is fairly simple to solve, and we obtain 
\be
h_\sc(T) = \frac{2G \rho_q}{\Lam - (3GT)^2}.
\ee
As in the preceding example, the Universe contracts, bounces at $T=0$ and expands again. However, we see here that $h_\sc \to \infty$ at a finite value of $T$ (a behavior that was previously noticed in~\cite{Roser14b}). This is simply due to the fact that the extrinsic curvature is given by $\dot a /a$, which is constant in a De Sitter space-time. This means that the Universe reaches a state of infinite volume in a finite York time. However, this should not be interpreted literally. As one can see from the results of section \ref{background_Sec}, any matter field will experience its \emph{proper time}, at least in the background field limit. In other words, it will oscillate an infinite number of times before $h$ reaches $\infty$. 

\begin{figure}[!ht]
\begin{center}
\includegraphics[width=0.8\columnwidth]{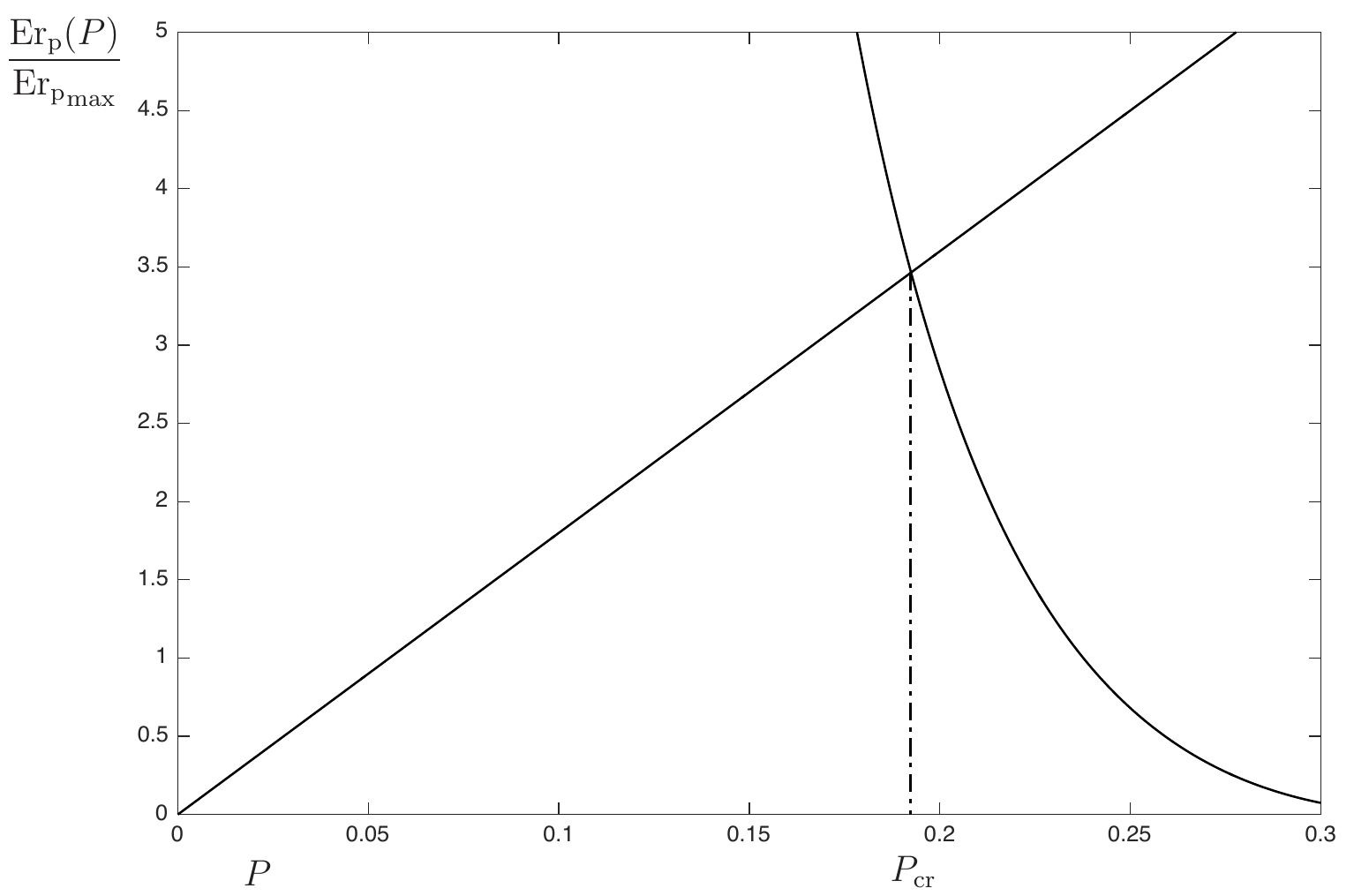}
\end{center}
\caption{Plot of the error functions $\Erp/\mathrm{Er}_{\rm max}$ and $\Erx/\mathrm{Er}_{\rm max}$ as a function of $P = GT/\Lam^{1/2}$. As in Fig.~\ref{Error_plot}, the functions are symmetric about $P=0$, so we only plotted $P>0$.}
\label{Fig_cross} 
\end{figure}

We now analyze the validity regime of the semiclassical approximation. Near the turning point ($h_\sc = 2G \rho_q/\Lam$), the momentum representation is accurate. When increasing $T^2$, $\Erp$ grows until the approximation breaks down. On the contrary, $\Erx$ is very large close to the turning point, but decreases with increasing values of $T^2$. At some value $T_{\rm cr}$, or equivalently, $h_{\rm cr} = h_\sc(T_{\rm cr})$, the two error bounds cross (see Fig.~\ref{Fig_cross}). On one side of this crossing, the momentum representation offers the best approximation, on the other side, the metric one is more accurate. 

To obtain a globally defined solution, valid for all values of $T$, we construct a combination of both representations, so as to exploit the most accurate one in each range of $T$. For this, we build 
\be \label{crossingWKB}
\tilde{\Psi}(T) = \Theta(T^2-T_{\rm cr}^2) \int {\chi}(h) e^{-i h T} \frac{dh}{\sqrt{2\pi}} + \Theta(T_{\rm cr}^2 - T^2) \tilde{\chi}(T), 
\ee
where $\tilde \chi$ is given by \eq{pWKB} and $\chi$ by \eq{geoWKB}. Now two remarks are in order. First, by construction, the error made in this approximate solution is bounded by the minimum of the error functions \eqref{WKBvalid} and \eqref{pWKBvalid}. A short computation shows that $\mathrm{Er}_{\rm max} \propto \frac{\Lam^{1/2}}{\V \rho_q}$. When $\mathrm{Er}_{\rm max} \ll 1$, \eq{crossingWKB} is a good approximation everywhere, i.e. close to the turning point and asymptotically. Second, from the results of Sec.~\ref{background_Sec}, we conclude that, up to $O(\mathrm{Er}_{\rm max})$, $\tilde \Psi$ is smooth, and the precise value of $T_{\rm cr}$ does not matter. Indeed, both expressions coincide when both error terms are small, which is the case in a vicinity of $T_{\rm cr}$.

It is noticeable that the maximum error obtained using \eq{crossingWKB} coincides with the one of Sec.~\ref{Solvable_ex}. This suggests that there exists a more general framework, allowing us to obtain semiclassical states in a representation independent manner, i.e. irrespectively of $a(h)$ or momentum \emph{vs.} metric representation. The natural direction to obtain such a framework would be to use a covariant approach, i.e. path integral methods. Unfortunately, if semiclassical solutions can be obtained via saddle point methods, to obtain precise characterizations of their validity is a very difficult task. In molecular physics, where similar questions arise, this point seems to remain unclear (see~\cite{Pechukas69b} and the discussion in Sec.~IV~A of~\cite{Delos72II}).

\subsection{Tunneling amplitude from the momentum representation}
\label{Tunnel_Sec}
\begin{figure}[!ht]
\begin{center}
\includegraphics[width=0.45\columnwidth]{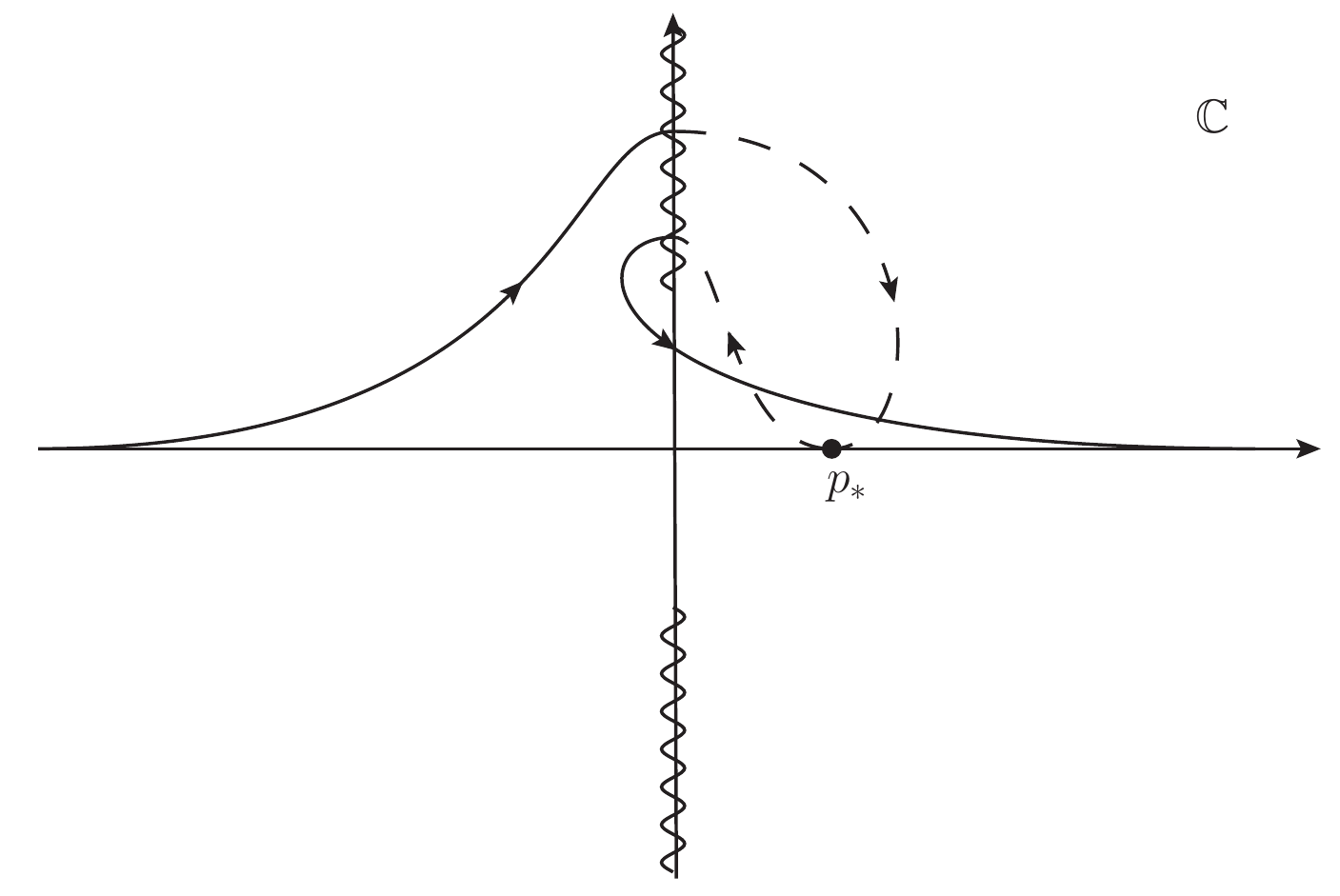}
\includegraphics[width=0.45\columnwidth]{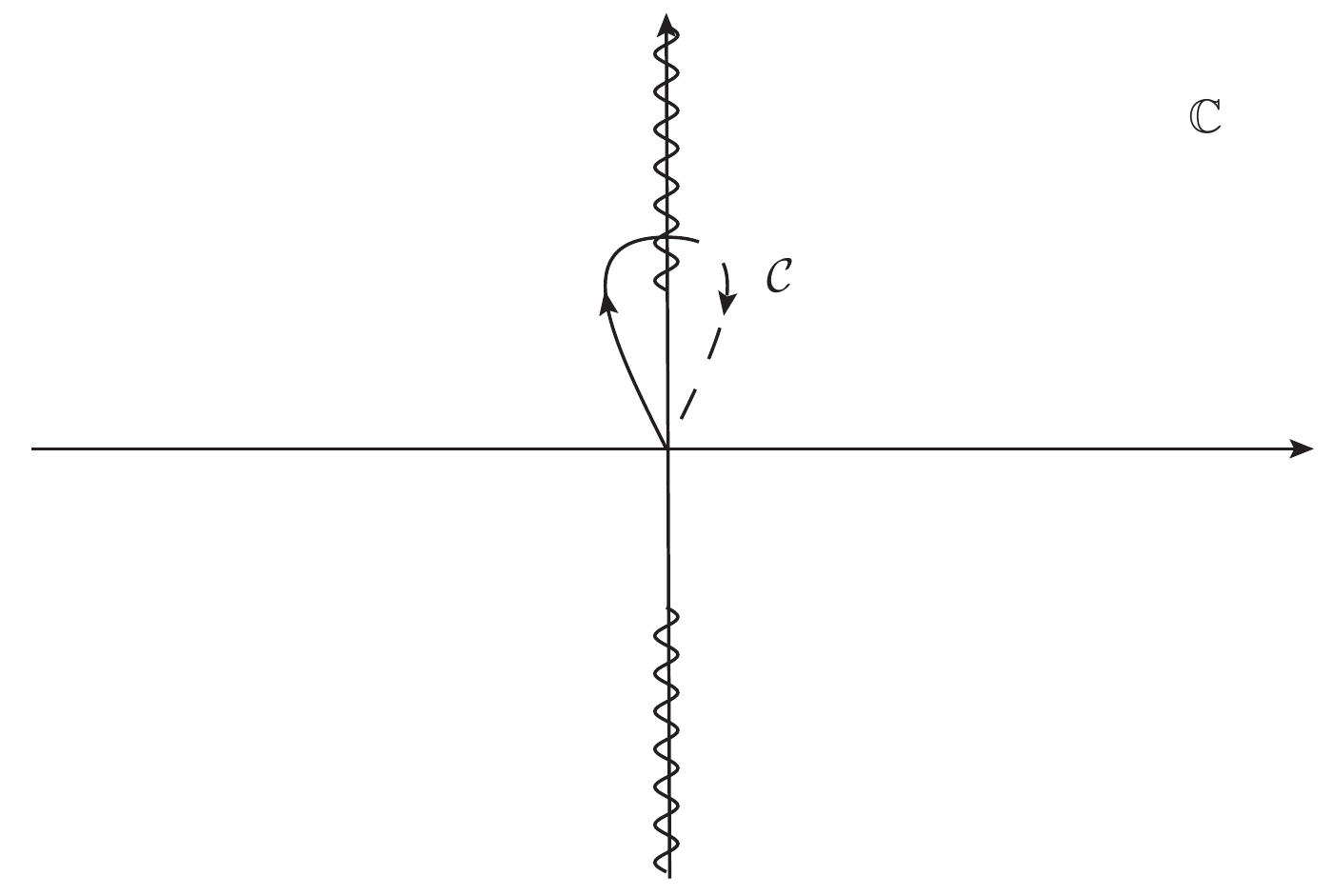}
\end{center}
\caption{Contours in the complex plane of $p_h$. On the left: the real line is deformed into the represented contour, in order to evaluate \eqref{Tunnel_pWKB} with the saddle point method. The new contour goes across the branch cut to pick up the saddle point $p_*$. On the right: contour going from 0 to 0 on the other Riemann sheet, and used to obtain the tunnel amplitude in \eq{Tunnel_ampl}. 
}
\label{Contour_Fig} 
\end{figure}

As we have seen in Sec.~\ref{Solvable_ex}, it is not necessary to go back to the geometric representation. Depending on the problem at hand, one might want to stay in the momentum representation all along, so as to use $p_h$ as a time variable to parametrize the rate of events (such as particle production~\cite{Coutant14}). However, in certain cases, going back to the geometric representation from the momentum allows one to go beyond the semiclassical approximation. This is the case when there are two distinct classically forbidden regions, and the Universe can tunnel from one to another. As we now show, the integral expression obtained from the momentum representation, i.e. \eq{inverseFT}, naturally gives the tunneling amplitude. 
To illustrate this, we shall investigate in detail the simplest potential barrier: the harmonic one. We consider the Wheeler-DeWitt equation \eqref{Schro} with the potential 
\be
W(h) = W_{\rm max} - \lam h^2. 
\ee
Using this expression in the Hamilton-Jacobi equation \eqref{pHJ}, the solutions read  
\be \label{Tunnel_ptraj}
h_\sc(p_h) = \pm \sqrt{\frac{W_{\rm max} + p_h^2}\lam}. 
\ee
By picking a sign, one selects one semiclassical branch, living either on the left or right side of the turning points $h = \pm h_\tp = \pm \sqrt{W_{\rm max}/\lam}$. Without loss of generality, we shall choose the left branch. As we shall see, the Fourier transform of momentum semiclassical solutions gives us the amplitude to tunnel on the other side. From \eq{Tunnel_ptraj} we directly deduce the momentum semiclassical solution \eqref{pWKB}
\be
\tilde \Psi(p_h) = \frac{\exp\left(i \lam^{-1/2} \int^{p_h}_{p_0} \sqrt{W_{\rm max} + p^2} dp\right)}{\left(4\lam (W_{\rm max} + p_h^2)\right)^{1/4}}, 
\ee
where $p_0$ is a conventional reference momentum value, chosen to be 0 for convenience. We can now build the inverse Fourier transform as a solution in geometric representation 
\be \label{Tunnel_pWKB}
\Psi(h) = \int \frac{\exp\left(i p_h h + i \lam^{-1/2} \int^{p_h}_{0} \sqrt{W_{\rm max} + p^2} dp\right)}{\left(4\lam (W_{\rm max} + p_h^2)\right)^{1/4}} dp_h. 
\ee
We directly see that when $h < -h_\tp$, we may use the saddle point approximation directly as in Sec.~\ref{invFT_Sec}, and obtain a semiclassical solution reflected on the potential barrier (as in \eq{spReflection}). For $h > h_\tp$, the wave function in \eq{Tunnel_pWKB} is \emph{a priori} nonzero. In fact, its asymptotic for $h \to +\infty$ gives the amplitude of the wave that has tunneled across the potential barrier. However, this asymptotic is slightly delicate to obtain, as the saddle point equation seems to have no solution for $h > h_\tp$. In fact, if we analytically continue $h_\sc(p_h)$, the saddle point $p_*$ lies \emph{on the other side of the branch cut} of the square root. Therefore, to pick it up, one must first deform the contour across the branch cut so as to go to the second Riemann sheet, where the square root flips signs (see Fig.~\ref{Contour_Fig}). Hence, we use the identity 
\be
\int^{p_*}_{0_1} \sqrt{W_{\rm max} + p^2} dp = \int_{\mathcal C} \sqrt[\mathbb C]{W_{\rm max} + p^2} dp - \int^{p_*}_{0_2} \sqrt{W_{\rm max} + p^2} dp, 
\ee
where the contour $\mathcal C$ is defined in Fig.~\ref{Contour_Fig}. We can now use the saddle point method to evaluate \eqref{Tunnel_pWKB} and we find 
\be
\Psi(h \gg h_\tp) \sim A_T \frac{e^{i \int^h p_\sc(h') dh'}}{\sqrt{|2p_\sc(h)|}}, 
\ee
where $p_\sc(h) = \sqrt{\lam h^2 - W_{\rm max}}$. This describes the part of the wave function that has tunneled across the potential barrier. $A_T$ is the tunneling amplitude given by 
\be \label{Tunnel_ampl}
A_T = e^{i \lam^{-1/2} \int_{\mathcal C} \sqrt[\mathbb C]{W_{\rm max} + p^2} dp}. 
\ee
The tunneling amplitude is thus governed by the contour integral $\int_{\mathcal C} \sqrt[\mathbb C]{W_{\rm max} + p^2} dp$. A short calculation then shows  
\bsub
\int_{\mathcal C} \sqrt[\mathbb C]{W_{\rm max} + p^2} dp &=& 2W_{\rm max} \int_{0}^{i} \sqrt[\mathbb C]{1 + t^2} dt, \\
&=& i \frac{\pi}4 W_{\rm max}. 
\esub 
Hence, the tunneling amplitude reads 
\be
|A_T|^2 = e^{- \pi W_{\rm max}/\sqrt{\lam}}. 
\ee
This is the same result as one would have obtained by working directly in the geometric representation and using connection formulas. More generally, the fact that analytic continuations of semiclassical solutions give tunneling amplitudes is a well-known feature of semiclassical methods~\cite{Berry72}, but to our knowledge, it has never been obtained from semiclassical solutions in momentum representation. This provides an alternative derivation. The interest is that one can keep working in momentum representation to investigate ``local'' phenomena, such as particle creation near the turning point~\cite{Coutant14}, but the solution will still keep track of the tunneling across the potential barrier, and the amount of tunneling can be obtained \emph{in fine}. 

\section{Conclusion}

In this work, we analyzed the general framework of~\cite{Coutant14}, which gives semiclassical solutions of the Wheeler-DeWitt equation in the momentum representation. The aim was to establish the key advantages of working in momentum representation. First, it provided solutions whose domain of validity are orthogonal to the standard WKB approximation, and in particular, are valid around turning points. This was shown by providing a bound for the error made by our semiclassical approximation and comparing it with the WKB one, see Eqs.~\eqref{pWKBvalid} and~\eqref{WKBvalid}. When both approximations hold, we showed that they give equivalent results (see Sec.~\ref{invFT_Sec}). Second, one needs not to go back in geometric representation. Indeed, when \eqref{pWKBvalid} is satisfied, the momentum semiclassical solutions not only provide accurate solutions, but the momentum becomes a good time variable, that can be used to investigate the effective dynamics of matter, something impossible with other turning point resolutions based on Airy functions. Third, despite its semiclassical character, a proper analytic continuation of these solutions gives access to nonclassical quantity, such as tunneling amplitudes (see Sec.~\ref{Tunnel_Sec}). Fourth, we presented this method in a very general framework (for any choice of ordering of the Wheeler-DeWitt constraint, and any choice of parametrization of the metric, see Sec.~\ref{pWKB_Sec}), allowing for numerous applications. In particular, in Sec.~\ref{York_ex}, we showed how to apply our framework to the use of York time starting from the Wheeler-DeWitt equation. Unlike in the ``choosing time before quantization'' approach~\cite{Isham92}, where one first solves the Hamiltonian constraint for the desired variable and then quantizes, we control here under which conditions this \emph{ad hoc} procedure is a good approximation to the general Wheeler-DeWitt equation. Since York time is the conjugate momentum to the volume element~\cite{York73}, our framework directly gives semiclassical solutions parametrized by York time, and allows to control their validity regime. The study of our examples also hints toward the existence of a unified framework, allowing us to minimize the error of the semiclassical approximation. We also believe our results are directly applicable to loop quantum cosmology, in the so-called $b$-representation~\cite{Ashtekar07}.

\acknowledgements
I thank Casey Tomlin and Edward Wilson-Ewing for useful the comments on this manuscript.

\newpage
\appendix
\section{Gradient expansions}
\label{pWKB_App}

\subsection{Noncommutative Taylor expansion}

In this appendix, we shall give a proof of the noncommutative expansion used in Sec.~\ref{pWKB_Sec}. (To simplify, we drop the indices used in the core of the text, i.e. $p_h = p$ and $h_\sc(p) = h(p)$.) The aim is to expand the operator $W(h(p)+ i\p_p)$ in powers of $i\p_p$. Note that this operator is perfectly well defined. It can be built rigorously as the composition of the multiplication operator $M_W: \psi(z) \to W(z) \psi(z)$, with the Fourier operator $\mathcal F$ and the (unitary) multiplication operator $M_{\exp}: \psi(z) \to \exp(-i \int h(z)dz) \psi(z)$. We then define 
\be \label{MathDef}
W(h(p)+ i\p_p) = M_{\exp}^\dagger \mathcal F M_W \mathcal F^\dagger M_{\exp} . 
\ee
To obtain its Taylor expansion, we assume that the function $W(z)$ is analytic, and possesses the following series representation 
\be
W(z) = \sum_{n \in \mathbb N} \alpha_n z^n. 
\ee
The operator we wish to expand is therefore given by 
\be \label{TaylorFull}
W(h(p)+ i\p_p) = \sum_{n \in \mathbb N} \alpha_n \left(h(p) + i \p_p \right)^n.
\ee
Expanding this equation in powers of $i\p_p$ is a well defined procedure. By expanding the series term by term, we shall see that it is rather easy to sort out the various powers of $i\p_p$. The delicate point however, is to understand the validity of a \emph{truncation} of the series, as being a ``good approximation'' of the full operator. In the next subsection, we return to that point and characterize the error made by such a truncation using a next-to-leading order computation. Back to \eq{TaylorFull}, its $n$-th term gives 
\be
\left(h(p) + i \p_p \right)^n = h(p)^n + \sum_{m=0}^{n-1} h(p)^{n-1-m} i \p_p h(p)^{m} + O(\p_p^2). 
\ee
The second term contains all the possible orderings of $i\p_p$ and $h(p)$. To simplify it, we notice that 
\bsub
\sum_{m=0}^{n-1} h(p)^m i \p_p h(p)^{n-1-m} &=& \sum_{m=0}^{n-1} \left(h(p)^{n-1} i \p_p + i m h(p)^{n-2} h'(p) \right) , \\
&=& n h(p)^{n-1} i\p_p + \frac i2 n(n-1) h(p)^{n-2} h'(p), 
\esub
where the well-known identity $\sum_{m=1}^{n-1} m = n(n-1)/2$ has been used. Plugging this result into \eq{TaylorFull} and summing up, we directly obtain 
\be
W(h(p)+ i\p_p) = W(h(p)) + W'(h(p)) i\p_p + \frac i2 W''(h(p)) h'(p) + O(\p_p^2). 
\ee
This can also be written in a more symmetric way as 
\be \label{Wexp1st}
W(h(p)+ i\p_p) = W(h(p)) + \frac 12 \big( W'(h(p)) i\p_p + i\p_p W'(h(p)) \big) + O(\p_p^2). 
\ee
This form shows that the first order term is simply the only ordering that preserves the Hermitian character of the operator $W(h(p)+ i\p_p)$. 

\subsection{Next-to-leading order and error function}
By using the same method as before, a lengthy but straightforward calculation gives the second order expansion of \eq{Wexp1st}
\be
O(\p_p^2) = \frac14 i\p_p W''(h(p)) i\p_p + \frac18 \left((i\p_p)^2 W''(h(p)) + W''(h(p)) (i\p_p)^2\right) + O(\p_p^3). \label{ordering}
\ee
From this, one can obtain an estimation of the error due to the semiclassical approximation~\footnote{We present here an improved version of the proof of appendix C in~\cite{Coutant14}.}, that is, the validity condition \eqref{pWKBvalid}. For this, we decompose a solution of the Wheeler-DeWitt equation \eqref{pSchro} as 
\be
\tilde \Psi(p) = \tilde \chi(p)(1+\eps(p)). 
\ee
$\tilde \chi$ is the semiclassical solution \eqref{pWKB}, and $\eps$ encodes the corrections. A next-to-leading order gradient expansion of the Wheeler-DeWitt equation \eqref{pSchro} gives an equation on the error function. 
\be
W' A(p) i \eps'(p) + \frac14 i\p_p W'' i\p_p A(p) + \frac18 \left((i\p_p)^2 W'' + W'' (i\p_p)^2\right)A(p) = 0, 
\ee
where $A$ is the semiclassical amplitude of \eq{pWKB} and $W$ and its derivatives are evaluated on $h=h(p)$. To ease the discussion, we assume $W'>0$ and since $A(p) = W'(p)^{-1/2}$, we directly obtain $\eps'(p)$ in terms of $W$ and its derivatives. 
\be
\eps'(p) = i \frac{48 W' W'' W''' h'^2 + 6 W' W'' h'' - 9 W''^3 h'^2 - 4 W^{(4)} W'^2}{32W'^3}. \label{epsprime}
\ee
Although $\eps$ is a local function of $p$, what is relevant to characterize the semiclassical approximation is the \emph{accumulated error}. Starting at a point $p=p_0$, $\chi$ always matches an exact solution of the Wheeler-DeWitt equation (with appropriate initial conditions). At a later point $p_1$, the relative error is then bounded by $\int_{p_0}^{p_1} |\eps'(p)| dp$ (see for instance~\cite{Winitzki05,Olver} for the standard WKB approximation, and the appendix of~\cite{Coutant11} in a context of higher order differential equations). If the variations of $\eps'$ are smooth enough, this integral is bounded by the maximum value of $\eps$. In addition, if the derivatives of the potential are well sorted, all terms in \eq{epsprime} contribute to the same order. The numerical factors of the various terms above are thus irrelevant, and one can tune them to integrate $\eps'(p)$ in a simple manner. Doing so, we obtain the estimate for the error function as  
\be \label{eps_brut}
\eps(p) \simeq i \frac{W''^2 h'}{W'^2} .
\ee
The criterion of \eq{pWKBvalid} is then simply $|\eps| \ll 1$. Note that this condition is very general. In fact it would still apply if the constraint were of the form $Q(p) + W(h) = 0$ rather than $p^2 + W(h)$. In our case, one can use the the Hamilton-Jacobi equation $p^2 + W(h(p)) = 0$ to rewrite \eqref{eps_brut} and eliminate $h'$. This gives alternative forms 
\be 
\eps(p) \simeq - i \frac{2W''^2 p}{W'^3} \simeq - i \frac{2W''^2 W^{1/2}}{W'^3} .
\ee
The last form presents the advantage to expressing the error function only in terms of $W$ and its derivatives.

\bibliographystyle{utphys}
\bibliography{Bibli}

\end{document}